\documentclass[conference]{IEEEtran}
\IEEEoverridecommandlockouts
\usepackage{cite}
\usepackage{amsmath,amssymb,amsfonts}
\usepackage{algorithmic}
\usepackage{graphicx}
\usepackage{textcomp}
\usepackage{xcolor}
\def\BibTeX{{\rm B\kern-.05em{\sc i\kern-.025em b}\kern-.08em
    T\kern-.1667em\lower.7ex\hbox{E}\kern-.125emX}}
\begin{document}

\title{Finding important edges in networks through local information\\
}

\author{\IEEEauthorblockN{En-Yu Yu}
\IEEEauthorblockA{\textit{school of computer science and engineering} \\
\textit{UESTC}\\ \\
943530714@qq.com}
\and
\IEEEauthorblockN{Yan Fu}
	\IEEEauthorblockA{\textit{school of computer science and engineering} \\
		\textit{UESTC}\\
		fuyan@uestc.edu.cn}
\and
\IEEEauthorblockN{Jun-Lin Zhou}
	\IEEEauthorblockA{\textit{school of computer science and engineering} \\
		\textit{UESTC} \\
		jlzhou@uestc.edu.cn}
\and
\IEEEauthorblockN{Duan-Bing Chen*}
	\IEEEauthorblockA{\textit{school of computer science and engineering} \\
		\textit{UESTC} \\
			\textit{Union Big Data Tech. Inc.} \\
		dbchen@uestc.edu.cn(Correspondence)}
}

\maketitle

\begin{abstract}
In transportation, communication, social and other real complex networks, some critical edges act a pivotal part in controlling the flow of information and maintaining the integrity of the structure. Due to the importance of critical edges in theoretical studies and practical applications, the identification of critical edges gradually become a hot topic in current researches. Considering the overlap of communities in the neighborhood of edges, a novel and effective metric named subgraph overlap (SO) is proposed to quantifying the significance of edges. The experimental results show that SO outperforms all benchmarks in identifying critical edges which are crucial in maintaining the integrity of the structure and functions of networks.
\end{abstract}


\section{Introduction}
With the acceleration of global informatization, human life is closely related to various complex networks\cite{albert2002statistical, costa2011analyzing, gao2019computational, wang2019coevolution}. Electricity, water and gas networks affect people's daily life\cite{watts1998collective}; road, railway and aviation networks affect people's travel\cite{ghosh2011statistical}; various popular social networks affect the spiritual life of individuals and the entire society\cite{weng2010twitterrank}. In real networks, a few nodes and edges act pivotal roles and have great influence on the structure and functions of networks\cite{caldarelli2007scale, li2017simple, liao2017information, zhou2018overlapping}. Identifying critical nodes and edges has widely used in many applications such as analysis of cascading failures, control of infectious diseases and marketing of goods. In previous researches, the significance of nodes has been measured by numerous methods\cite{lu2016vital, chen2012identifying, yu2020identifying, chen2019identifying, yu2020re}, yet how to evaluate the importance of edges receives less attention. In a complex network, the scale of edges is larger than that of nodes and the complexity of  networks is often determined by edges. Therefore, the identification of critical edges is more difficult and meaningful\cite{holme2002attack, xia2008attack}.

To identify critical edges, current methods mainly focus on the structural information of networks. Ball et al. \cite{ball1989finding} pointed out that the importance of an edge can be measured by the change of the average distance of the network after removing this edge. Similar to the betweenness centrality of nodes\cite{freeman1977set}, Newman et al.\cite{girvan2002community} used the betweenness of edges (EB) to quantify the importance of edges. Yu et al.\cite{yu2018identifying} proposed a method named $BCC_{MOD}$, and it was significantly better than EB on all test networks. These algorithms based on global information have good results on small-scale networks, however, they are unsuitable for large-scale networks since they are time-consuming. Many researchers begin to use local information to characterize the significance of edges in order to reduce the time complexity. Holme et al.\cite{holme2002attack} supposed that edges between critical nodes are more important than other edges and proposed degree product (DP) index to evaluate the significance of edges. Consider the influence of node's common neighbors on the importance of edges, topological overlap (TO) was proposed by Onnela et al.\cite{onnela2007structure}. Cheng et al.\cite{cheng2010bridgeness} found that edges in a small clique with its endpoints in large cliques are important in connecting the network. Based on this idea, the bridgeness (BN) index was proposed. Liu et al.\cite{liu2015improving} proposed diffusion intensity (DI) to identify critical edges from the perspective of spreading dynamics. Besides, there are many other methods such as eigenvalues\cite{restrepo2006characterizing}, link entropy\cite{qian2017quantifying} and nearest neighbor connections\cite{ouyang2018quantifying} to measure the importance of edges, which will not be introduced here. 

In this paper, considering the overlap of communities in the neighborhood of edges, a novel and effective index named subgraph overlap (SO) is proposed. In SO index, the importance of an edge is characterized by the overlap of communities in its second-order neighborhood. In the experimental section, the performance of SO and five benchmark methods (DP, TO, DI, BN, SN) is measured by the robustness index\cite{schneider2011mitigation}. The results on different networks show that SO is better than other methods in identifying critical edges which are crucial in maintaining the communication among different communities in networks. In addition, as a local index,SO is suitable for large-scale networks.

\section{Theory and Methods}
\subsection{Network}
\begin{figure*}[htbp]
	\centerline{\includegraphics[width=14cm, height=5cm]{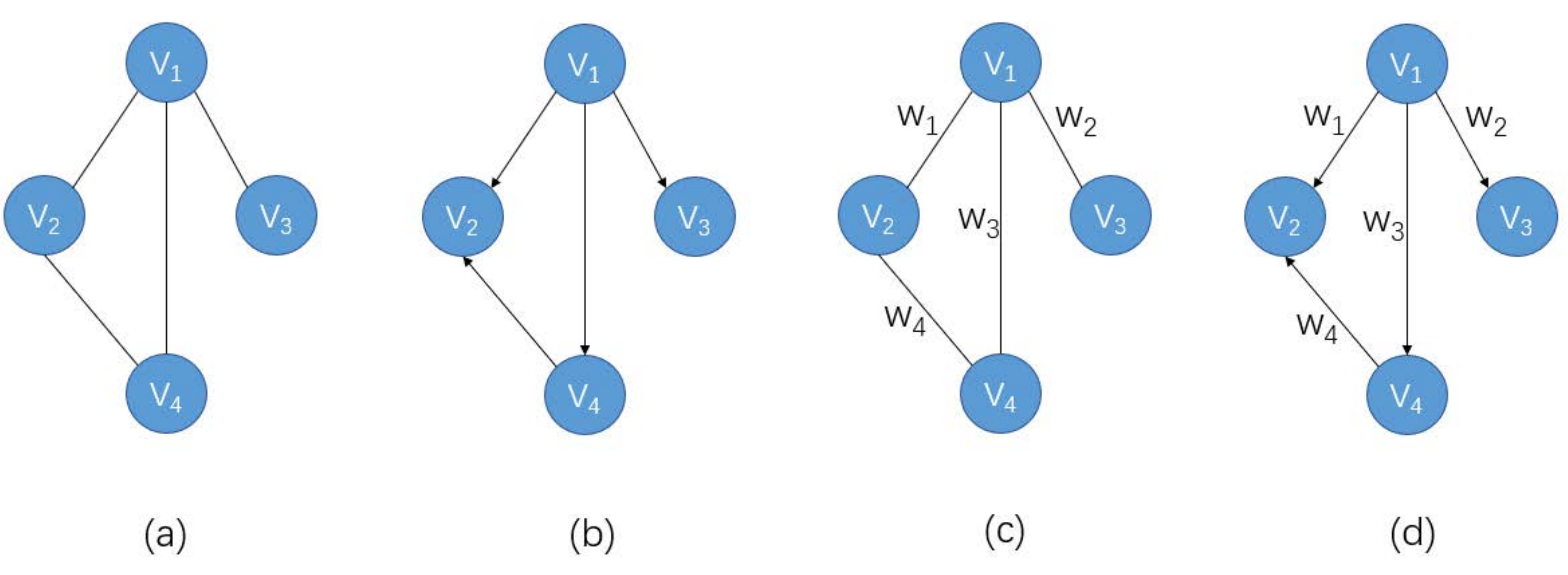}}
	\caption{Schematic diagram of static network: (a) undirected, unweighted; (b) directed, unweighted; (c) undirected, weighted; (d)directed, weighted. }
	\label{fig:static}
\end{figure*}
A static network is usually expressed as $G=(V, E)$, where $V=\{v_1,v_2,\dots,v_{N}\}$ is the node set and $E\subseteq V\times V$ is the edge set ($|V|=N, |E|=M$). Usually, we use $e_{ij }$ to represent the edge connecting node $i$ and node $j$. And the adjacency matrix $\mathrm {A}\in \mathbb{R}^{|V|\times |V|}$ is often used to calculate and store networks. The specific expression is as follows : 
\begin{equation}
\mathrm {A}[v_i, v_j]=a_{ij}=\begin{cases}
0 & \text{ if } (v_i,v_j)\notin E  \\
1 & \text{ if } (v_i,v_j)\in E
\end{cases}
\end{equation}
As shown in  Fig \ref{fig:static}, static networks contain static undirected networks, static directed networks, static weighted networks, and static unweighted networks according to whether the edges have directions and weights. In this paper, we are using unweighted networks.

In a static network $G$, The number of edges directly connected to node $i$ is often called the degree of node $i$ and expressed by $k_i$. For detail, $k_i$ is defined as: 
\begin{equation}
k_i=\sum_{j=1}^{N}a_{ij},
\end{equation}
and nodes directly connected to node $i$ are called neighbors of node $i$, denoted by $\Gamma_i$. $⟨k⟩$ is the average degree of $G$.

The distance between nodes is also one of the important structural parameters. In an static unweighted network, the path connecting node $i, j$ with the least edges is called the shortest path and the distance $d_{i, j}$ represents the number of edges included in it. In addition, the nodes whose distance to node $i$ is $l$ are called $l$-hop neighbors of node $i$.

\subsection{Subgraph Overlap Index}
Previous studies\cite{csermely2006weak, granovetter2018getting} pointed out that in facilitating communications among communities in the network, edges between two different communities in the network are often more important than those within communities. And edges between large communities are more important than other edges, and inversely proportional to other links between the two communities. At the same time, the method using global information is only can be used in small-scale networks due to the high time complexity, such as edge betweenness\cite{girvan2002community}. Inspired by above ideas, we tried to characterize the importance of edges by the overlap of communities in the neighborhood of edges. Then, a novel and effective index named subgraph overlap (SO) is proposed. 

\begin{figure}[htbp]
	\centerline{\includegraphics[width=8cm, height=4cm]{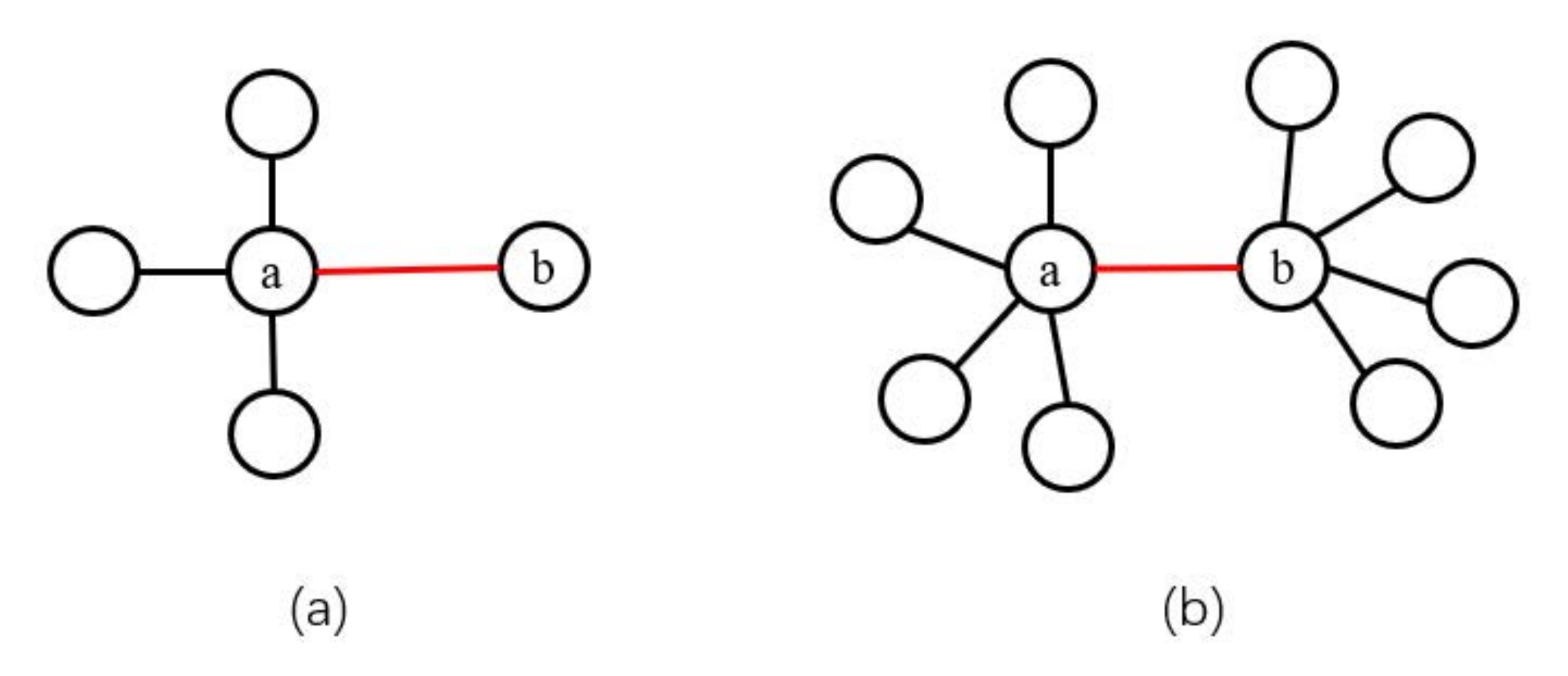}}
	\caption{Two toy subgraphs.}
	\label{fig:ex2}
\end{figure}

\begin{figure*}[htbp]
	\centering
	\includegraphics[width=16cm, height=6cm]{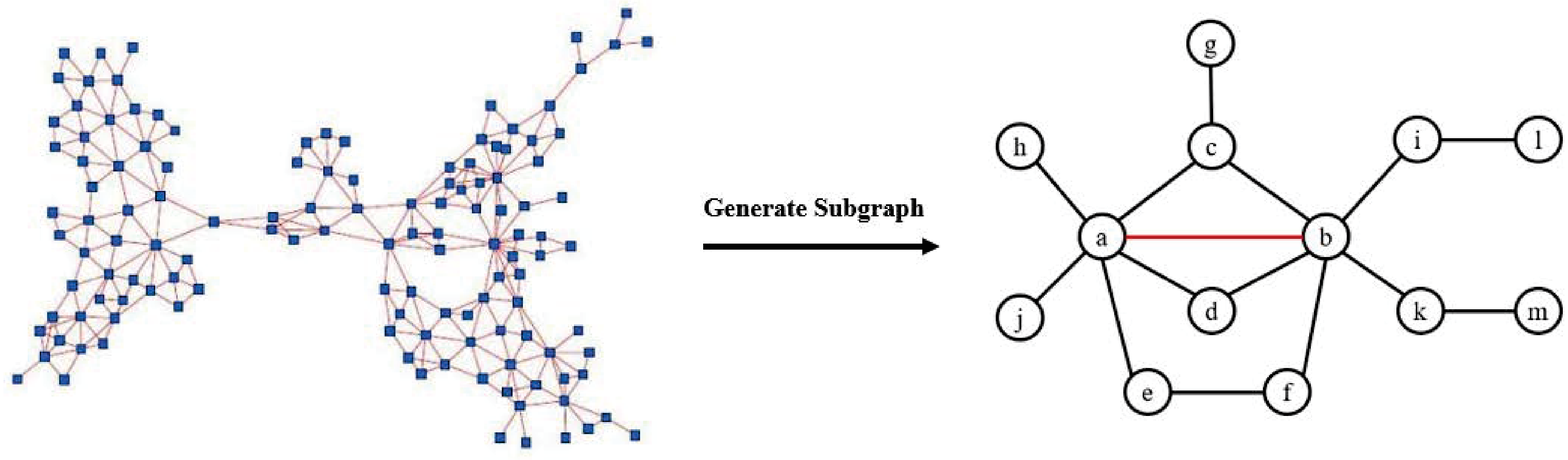}
	\caption{A simple example for SO index. The graph on the left is the entire network $G$. The graph on the right is a subgraph of $G$ which contains node $a, b$ and nodes whose distance to $a$ or $b$ is no more than 2. In the subgraph $G_{ab}$,  $\Gamma _a^{(2)}=\{v_a,v_b,v_c,v_d,v_e,v_g,v_f,v_h,v_j\}$ and $\Gamma _a^{(2)}=\{v_a,v_b,v_c,v_d,v_e,v_g,v_f,v_i,v_l,v_k,v_m\}$}
	\label{fig:example}
\end{figure*}

For a given static network $G$ and edge $e_{ij}$, $G_{ij}$ is a subgraph of $G$ which contains node $i, j$ and nodes whose distance to $i$ or $j$ is no more than 2. The SO index is defined as :
\begin{equation}
SO(i, j) = \frac{max(1, |\Gamma _i^{(2)} \cap \Gamma _j^{(2)}|)^2}{|G_{ij}|},
\end{equation}
where $|G_{ij}|$ represents the number of nodes in $G_{ij}$. $\Gamma _i^{(2)}$ is the set of nodes whose distance to node $i$ is no more than 2 in $G_{ij}\setminus {e_{ij}}$ ($\Gamma _i^{(2)}$ contains node $i$). Lower SO value indicates that the edge is more likely to connect large communities and there are few other connections between these two communities.

 In addition, as shown in Fig \ref{fig:ex2}(a) and (b), after removing $e_{ab}$, if the community where node $a$ is located is no longer connected to the community where node $b$ is located ($ |\Gamma _a^{(2)} \cap \Gamma _b^{(2)}|=0$), the importance of $e_{ab}$ should be measured by the size of subgraph $G_{ab}$, so the minimum value of the numerator is set to 1. The SO index only depends on local topological information and the time complexity is $O(M⟨k⟩^2)$. 

A brief introduction to the calculation process of SO index is shown in Fig \ref{fig:example}. In order to get $SO(a, b)$, we first extract the subgraph $G_{ab}$ from the original network, obviously, $|G_{ab}|=13$. Then we need to calculate $\Gamma _a^{(2)}$ and $\Gamma _b^{(2)}$ in $G_{ab}\setminus {e_{ab}}$. In the subgraph $G_{ab}$, $|\Gamma _a^{(2)} \cap \Gamma _b^{(2)}|=\{ v_a,v_b,v_c,v_d,v_e,v_g,v_f \}$, so $SO(a, b)=7^2/13=3.769$.

\begin{figure*}[htbp]
	\centerline{\includegraphics[width=16cm, height=16cm]{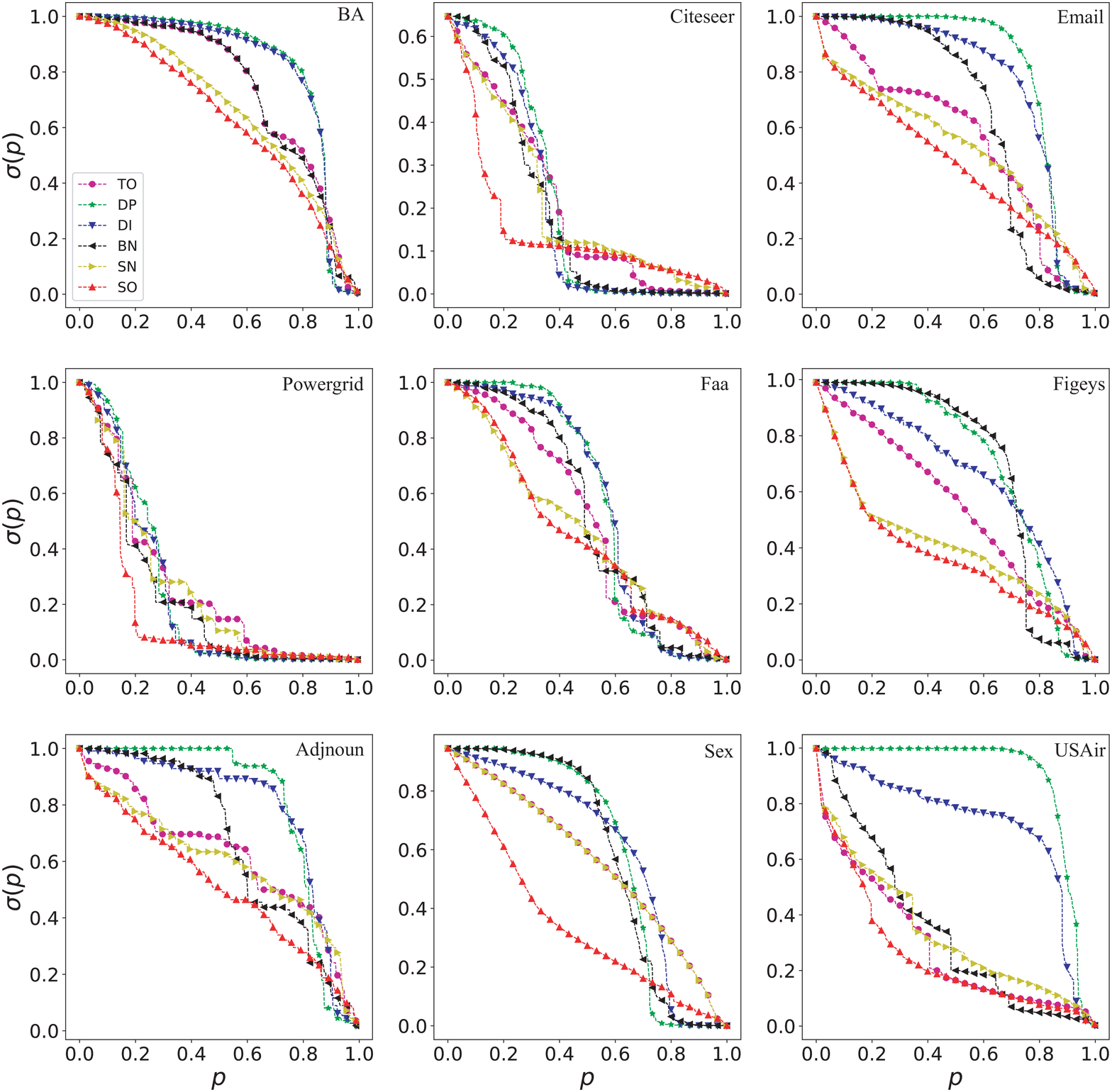}}
	\caption{The ratio of nodes of the largest connected component $\sigma$ after removing $p$ ratio of edges in nine networks.}
	\label{fig:R}
\end{figure*}
\begin{figure*}[htbp]
	\centerline{\includegraphics[width=16cm, height=16cm]{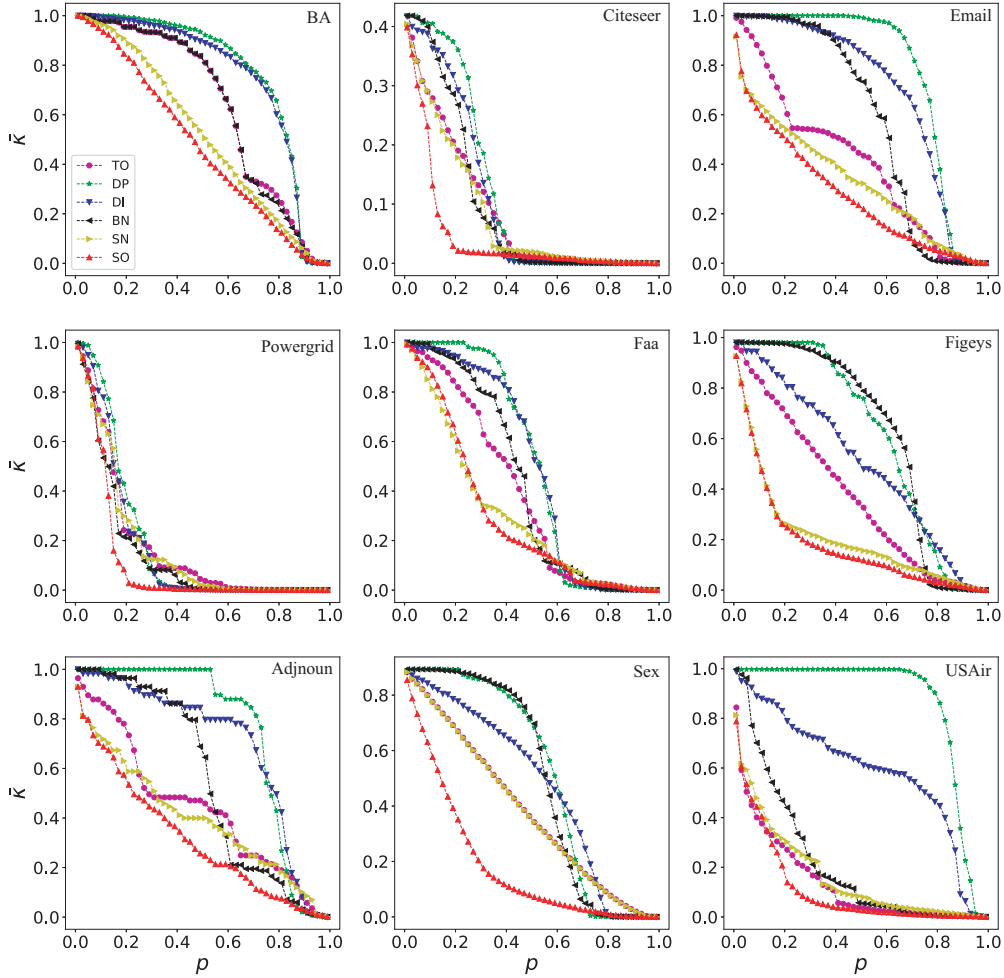}}
	\caption{With the removal of edges, the change of the average connectivity $\bar{\kappa}$.}
	\label{fig:C}
\end{figure*}
\begin{figure*}[htbp]
	\centerline{\includegraphics[width=16cm, height=16cm]{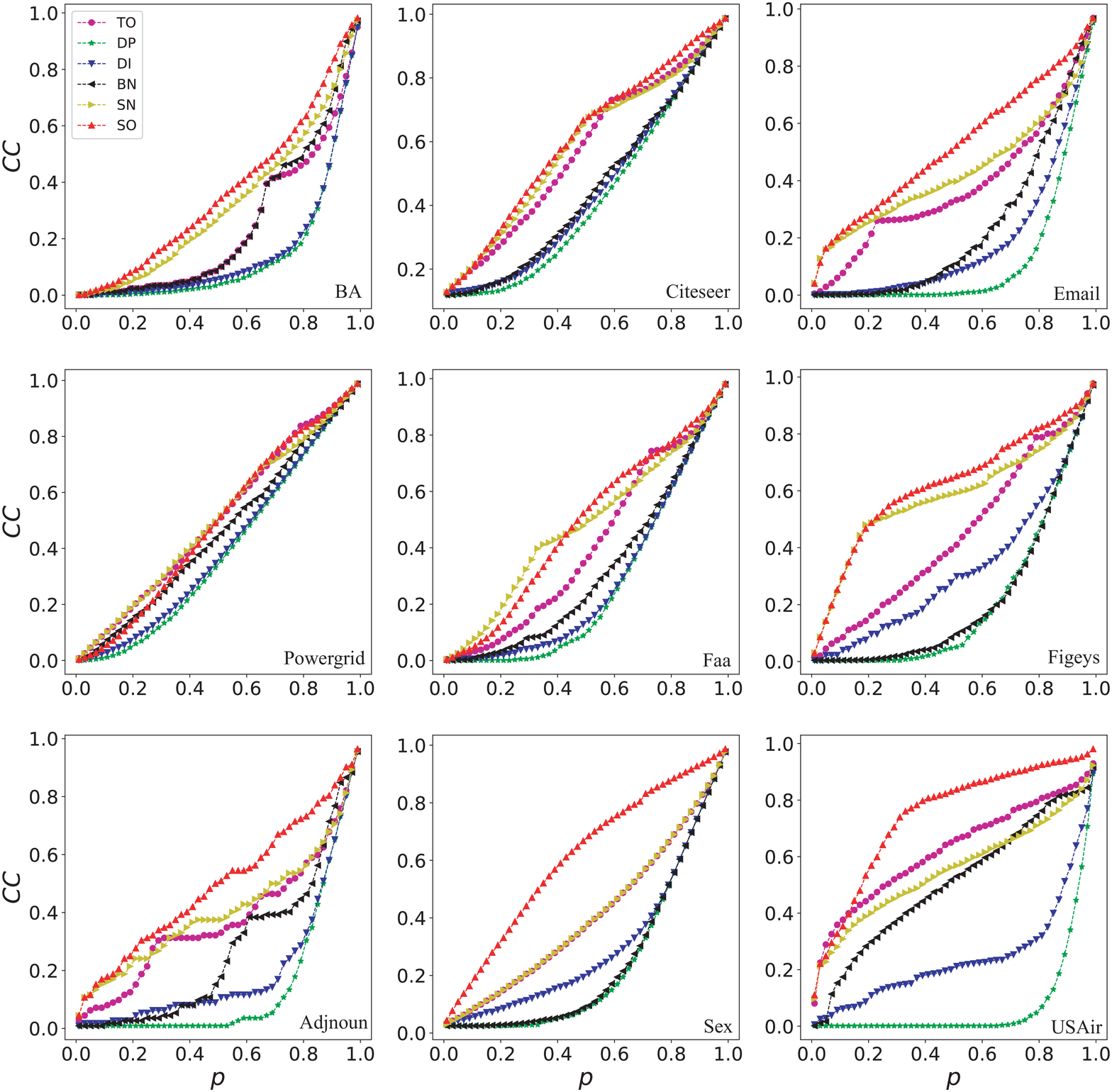}}
	\caption{The ratio of connected components $CC$ after removing $p$ ratio of edges in nine networks.}
	\label{fig:Co}
\end{figure*}

\subsection{Benchmark Methods}
In this paper, five well-known existing metrics such as topological overlap\cite{onnela2007structure}, degree product\cite{holme2002attack}, diffusion intensity\cite{liu2015improving}, bridgeness\cite{cheng2010bridgeness} and second-order neighborhood\cite{zhao2020identifying} are used to compare with SO. Same as SO, all benchmark methods use only local information. 

Degree product (DP)\cite{holme2002attack}:
\begin{equation}
DP(i, j) = k_ik_j.
\end{equation}

Topological overlap  (TO)\cite{onnela2007structure}:
\begin{equation}
TO(i, j) = \frac{|\Gamma_i \cap \Gamma_j|}{|\Gamma_i \cup \Gamma_j|-2},
\end{equation}
where $\Gamma _i$ is the set of neighbors of node $i$.

Diffusion intensity (DI)\cite{liu2015improving}:
\begin{equation}
DI(i, j) = \frac{(k_i-1)+(k_j-1)-2*|\Gamma_i \cap \Gamma_j|}{2}.
\end{equation}

Bridgeness (BN)\cite{cheng2010bridgeness}:
\begin{equation}
BN(i, j) = \frac{\sqrt{|S_i||S_j|}}{|S_{ij}|},
\end{equation}
where $S_i$ is the largest fully connected subgraph which contains node $i$ ($S_{ij}$ contains $e_{ij}$).

Second-order neighborhood (SN)\cite{zhao2020identifying}:
\begin{equation}
SN(i, j) =\frac{|s_i^{(2)} \cap s_j^{(2)} |}{|s_i^{(2)} \cup s_j^{(2)} |},
\end{equation}
where $s_i^{(2)}$ is the 2-hop neighbors of node $i$ in the subgraph $G\setminus {e_{ij}}$.

\section{Experiments}
\subsection{Datasets}

In experiments, the data set contains one synthetic and eight real networks. (1) BA, a synthetic network generated by the scale free model\cite{barabasi1999emergence}. (2)  Citeseer, a citation network contains a selection of the CiteSeer dataset\cite{sen2008collective}. (3) Email, an email network at an university\cite{konect:guimera03}. (4) Powergrid, a power network in the western America\cite{watts1998collective}. (5) Faa, an air traffic control network which is constructed from the USA's FAA \cite{konect}. (6) Figeys, a protein network\cite{konect:figeys}. (7) Adjnoun, an adjacent words network of the novel David Copperfield\cite{newman2006finding}. (8) Sex, a sexual intercourse network\cite{rocha2011simulated}. (9) USair, an airport transportation network in US\cite{konect}. Some basic topology features of each network can be found in Table \ref{tb.data}.
\begin{table}
	\centering
	\caption{The basic topology features. $\langle c \rangle$ and $H$ represent the average clustering coefficient and the degree heterogeneity, respectively.}\label{tb.data}
	\begin{tabular}{l llll ll}
		\hline
		Name&$N$&$M$&$\langle k\rangle $&$k_{max}$&$c$&$H$ \\ \hline
		BA     & 1000      &  4975     &  9.9500      & 112    &  0.0421  & 2.0808  \\
		Citeseer     & 3279      &  4552     &  2.7764     & 99    &  0.1435  & 2.4900  \\
		Email      & 1133      &  5451     &  9.6222       & 71    &  0.2201  & 1.9421  \\
		Powergrid      & 4941      &  6594     &  2.6690       & 19    &  0.0801  & 1.4503 \\
		Faa      & 1226      &  2408     &  3.9282    & 34    &  0.0675  & 1.8727  \\
		Figeys      & 2239      & 6432     &  5.7454   & 314    &  0.0399  & 9.7474  \\
		Adjnoun      & 112     & 425     &  7.5892    & 49   & 0.1728  & 1.8149  \\
		Sex      & 16730      &  39044     &  4.6675   & 305    &  0  & 6.0119  \\
		USair      & 1574      & 17215     & 21.8742       & 314    &  0.5042  & 5.1303  \\
		\hline
	\end{tabular}
\end{table}

\subsection{Results}

The performance of  SO is evaluated by edge percolation process\cite{callaway2000network, moore2000epidemics}. For detail, remove edges from the network in turn according to the ranking results of each method, after removing the same proportion of edges, the greater the change in the network, the more important the removed edges. In this paper,  the impact on the network connectivity after edges are removed is estimated by the famous measure named robustness $R$\cite{schneider2011mitigation} which is defined as
\begin{equation}
R=\frac{1}{M}\sum_{i=1}^{M}\sigma(i/M),
\end{equation}
where $\sigma(i/M)$ indicates the ratio of nodes in the maximum connected component after removing edges with the ratio $i/M$. Obviously, the method with smaller $R$ can decompose the network faster,  which means that it can better rank the edge significance.

Fig \ref{fig:R} shows the process of network decomposition in nine networks. It can be seen that SO is the fastest of all methods to reduces $\sigma$ to 0.2 in all networks except Faa. In Faa, SO is the fastest of all methods to reduces $\sigma$ to 0.4. The robustness $R$ of SO and other benchmark methods are shown in Table \ref{tb.R} and it is easy to find SO has the best result for each network. All results show that decomposing the network according to the results of SO can destroy the robustness of the network the fastest. This also proves that the ranking results given by SO are more reasonable.

\begin{table}
	\centering
	\caption{Comparison of $R$ in all networks with highlighting the best results in bold.}\label{tb.R}
	\begin{tabular}{l llll ll}
		\hline
		Networks&TO&DP&DI&BN&SN&SO\\ \hline
		BA      & 0.7419      &  0.8248     &    0.8143     & 0.7348    &  0.6633     & \textbf{0.6266} \\
		Citeseer    & 0.2041      &  0.2180     &0.1968     & 0.1920  &   0.2062     & \textbf{0.1472} \\
		Email      & 0.5534     & 0.8092     & 0.7623      & 0.6410    &  0.5233     & \textbf{0.4691} \\
		Powergrid     & 0.2651      &  0.2417    & 0.2320      & 0.2159  &    0.2567    & \textbf{0.1691}    \\
		Faa     & 0.5043     &  0.5636     &  0.5651       & 0.5238   &    0.4575     & \textbf{0.4454}   \\
		Figeys     & 0.5376     &  0.6999     &  0.6495       & 0.6847   &    0.4054     & \textbf{0.3691}  \\
		Adjnoun     & 0.6189      &  0.8005     &  0.7816       & 0.6655   &   0.5936    & \textbf{0.5114}   \\
		Sex     & 0.5568      &  0.5984     &  0.5965       & 0.5889    &    0.5568     & \textbf{0.3430}  \\
		USair     & 0.2851     &  0.8939     &  0.7244       & 0.3455  &    0.3342     & \textbf{0.2495}   \\
		\hline
	\end{tabular}
\end{table}

In addition to robustness $R$,  the impact on the network connectivity after edges are removed is also can be estimated by the average connectivity $\bar{\kappa}$:
\begin{equation}
\bar{\kappa}(G) = \frac{\sum_{u,v} \kappa_{G}(u,v)}{{N \choose 2}},
\end{equation}
 where ${N \choose 2}$ is the number of all possible node pairs in $G$. $\kappa_{G}(u,v)=1$ if node $u$ and node $v$ are reachable, otherwise $\kappa_{G}(u,v)=0$. In short, the average connectivity $\bar{\kappa}$ is the average node reachability in $G$. After removing a certain percentage of edges, the method with the smallest $\bar{\kappa}$ is the best. Fig \ref{fig:C} shows the change of  $\bar{\kappa}$ ($\bar{\kappa}(G\setminus{E_{remove}})$/$\bar{\kappa}(G)$) with edges being removed. The average $\bar{\kappa}(G)$ are shown in Table \ref{tb.C} and it is easy to see that SO has the best result for all networks. 
 
 \begin{table}
 	\centering
 	\caption{Comparison of average $\bar{\kappa}(G)$ (mean value of ordinates of all points in  Fig \ref{fig:C}) in all networks with highlighting the best results in bold.}\label{tb.C}
 	\begin{tabular}{l llll ll}
 		\hline
 		Networks&TO&DP&DI&BN&SN&SO\\ \hline
 		BA      & 0.6386     &  0.7733     &    0.7535     & 0.6327    &  0.5211     & \textbf{0.4744} \\
 		Citeseer    & 0.0866      & 0.1217     &0.1062     & 0.0966  &   0.0826     & \textbf{0.0457} \\
 		Email      & 0.4021     & 0.7750     & 0.6909      & 0.5729    &  0.3320     & \textbf{0.2770} \\
 		Powergrid     & 0.1779      &  0.1890    & 0.1713      & 0.1453  &    0.1721    & \textbf{0.1138}    \\
 		Faa     & 0.3878     &  0.5073     &  0.4924       & 0.4289   &    0.3007     & \textbf{0.2926}   \\
 		Figeys     & 0.3740     &  0.6140     &  0.5038       &0.6190   &    0.2088    & \textbf{0.1842}  \\
 		Adjnoun     & 0.4398      &  0.7540    & 0.6951    & 0.5515   &   0.3981   & \textbf{0.3173}   \\
 		Sex     & 0.3819      & 0.5161    &  0.4718       & 0.5001   &    0.3819     & \textbf{0.1853}  \\
 		USair     & 0.1369     &  0.8653    &  0.5901     & 0.2161  &  0.1654  & \textbf{0.1113}   \\
 		\hline
 	\end{tabular}
 \end{table}
 
 Fig \ref{fig:Co} shows the ratio of connected components $CC$ after removing $p$ ratio of edges in nine networks. $CC$ is defined as
 \begin{equation}
 CC(G) = \frac{|cc(G)|}{N},
 \end{equation}
 where $|cc(G)|$ is the number of connected components in $G$. $CC$ represents the degree of fragmentation of networks. After removing a certain percentage of edges, the method with the largest $CC$ is the best. The average $CC$ are shown in Table \ref{tb.Co} and it is easy to see that SO has the best result for all networks except Powergrid. In Powergrid, SO is very close to the best result. 
 
  \begin{table}
 	\centering
 	\caption{Comparison of average $CC$ (mean value of ordinates of all points in  Fig \ref{fig:Co}) in all networks with highlighting the best results in bold.}\label{tb.Co}
 	\begin{tabular}{l llll ll}
 		\hline
 		Networks&TO&DP&DI&BN&SN&SO\\ \hline
 		BA      & 0.2365    &  0.1417     &   0.1524     & 0.2527    &  0.3223    & \textbf{0.3699} \\
 		Citeseer    & 0.5700     &0.4231   &0.4435    & 0.4533  &   0.5838     & \textbf{0.6082} \\
 		Email      & 0.3815     & 0.1260     & 0.1900     & 0.2406    &  0.4304    & \textbf{0.5207} \\
 		Powergrid     & 0.5046      &  0.3910    & 0.4051     & 0.4590  &    \textbf{0.4989}    & 0.4911    \\
 		Faa     & 0.4121     &  0.2696    &  0.2841      & 0.3226   &   0.4782    & \textbf{0.4847}   \\
 		Figeys     & 0.4458     &  0.2144     &  0.3251       &0.2235   &    0.5807   & \textbf{0.6230}  \\
 		Adjnoun     & 0.3748      &  0.1489   & 0.1991   & 0.2742   &   0.4064  & \textbf{0.4964}   \\
 		Sex     & 0.4091     & 0.2477    &  0.3048      & 0.2538   &   0.4091  & \textbf{0.6109}  \\
 		USair     & 0.6220     &  0.0773   & 0.2436    & 0.5055 &  0.5526  & \textbf{0.7437}   \\
 		\hline
 	\end{tabular}
 \end{table}

\section{Conclusions}
 It is not an easy task to identify critical edges in various types of complex networks yet it is of both theoretical interests and practical importance. Through the study of communities overlap in the neighborhood of edges, a novel and effective index named subgraph overlap (SO) is proposed. SO can be used in large-scale networks with it only uses the local information of edges. The results of experiments on different networks show that SO is better than all benchmarks in identifying critical edges which are crucial in maintaining the communication among different communities in networks. SO have provided a new framework for quantifying the significance of edges. In the future work, this framework will be extended to temporal networks.

\section*{Acknowledgment}
This work is supported by the SSPP of UESTC under Grant No. Y03111023901014006, by NSFC under Grant No. 61673085.

\bibliographystyle{IEEEtran}
\bibliography{mybibfile}

\end{document}